\documentclass[conference]{IEEEtran}
\IEEEoverridecommandlockouts
\usepackage{cite}
\usepackage{amsmath,amssymb,amsfonts}
\usepackage{algorithmic}
\usepackage{graphicx}
\usepackage{textcomp}
\usepackage{xcolor}
\def\BibTeX{{\rm B\kern-.05em{\sc i\kern-.025em b}\kern-.08em
    T\kern-.1667em\lower.7ex\hbox{E}\kern-.125emX}}

\usepackage[utf8]{inputenc}
\usepackage{tikz}
\usetikzlibrary{shapes.geometric, arrows, positioning, matrix}

\usepackage{fancyhdr}
\usepackage[normalem]{ulem}
\usepackage[hyphens]{url}
\usepackage[final]{microtype}
\usepackage[nodisplayskipstretch]{setspace}
\usepackage{soul}
\sethlcolor{green}
\usepackage{comment}

\usepackage{graphicx}
\usepackage{textcomp}
\usepackage[binary-units=true]{siunitx}
\usepackage{mathtools}
\usepackage{booktabs}
\usepackage[ruled,vlined]{algorithm2e}
\usepackage{xcolor}
\usepackage{soul}

\usepackage{pifont}% http://ctan.org/pkg/pifont
\usepackage{tikz}

\newcommand{\cmark}{\ding{51}}%

\usepackage{enumitem}
\usepackage{siunitx}
\usepackage[ruled,vlined]{algorithm2e}
\usepackage{subcaption}
\usepackage{svg}
\svgpath{{../imgs/}}
\usepackage{enumitem}
\begin{document}
% \title{Swift: Accelerated Graph Analytics with Multiple FPGAs}
\title{Swift: A Multi-FPGA Framework for Scaling Up Accelerated Graph Analytics}
% \vspace{-1em}
%\title{The new document\vspace{-3em}}

%\begin{comment}
\author{\IEEEauthorblockN{
Oluwole Jaiyeoba\IEEEauthorrefmark{1}, 
Abdullah T. Mughrabi\IEEEauthorrefmark{2},
Morteza Baradaran\IEEEauthorrefmark{3},
Beenish Gul\IEEEauthorrefmark{4}, and 
Kevin Skadron\IEEEauthorrefmark{5}}
\IEEEauthorblockA{Department of Computer Science,
University of Virginia\\
Email: \IEEEauthorrefmark{1}oj2zf@virginia.edu,
\IEEEauthorrefmark{2}atmughra@virginia.edu,
\IEEEauthorrefmark{3}morteza@virginia.edu,
\IEEEauthorrefmark{4}bg9qq@virginia.edu,
\IEEEauthorrefmark{5}skadron@virginia.edu
}
}
%\end{comment}

\maketitle

\begin{abstract}
Graph analytics are vital in fields such as social networks, biomedical research, and graph neural networks (GNNs). However, traditional CPUs and GPUs struggle with the memory bottlenecks caused by large graph datasets and their fine-grained memory accesses. While specialized graph accelerators address these challenges, they often support only moderate-sized graphs (under 500 million edges). 
%Current scalable solutions rely on multiple machines with single FPGAs, which waste CPU resources, are energy inefficient, and add communication complexity. Scaling single-FPGA designs further introduces bottlenecks in PCIe communication, leading to idle FPGAs during graph iterations.

% Graph analytics have become essential across a wide range of application domains, such as social network analysis, biomedical research, and graph neural networks (GNNs). However, the widespread adoption, complexity, and increasing size of modern graphs present significant challenges for traditional CPUs and GPUs due to the high volume of random, fine-grained memory accesses caused by large graph datasets, which create a memory bottleneck. Specialized hardware graph accelerators have been developed to address these issues, but their capabilities are often limited to graphs of moderate size (less than 500 million edges). Existing accelerated solutions that can be deployed today typically use a single Field Programmable Gate Arrays (FPGA), and to scale up to large graphs, spreading the workload across multiple machines (each with a single FPGA). Using multiple compute nodes wastes CPU resources and is hugely energy inefficient, and suffers  added complexity in managing node-to-node communication. Moreover, when scaling up a single-FPGA design, the PCIe communication channel among FPGAs becomes a bottleneck and causes FPGAs to remain idle during message exchange at the end of each graph iteration.
Our paper proposes Swift, a novel scale-up graph accelerator framework that processes large graphs by leveraging the flexibility of FPGA custom datapath and memory resources, and optimizes utilization of high-bandwidth 3D memory (HBM). Swift supports up to 8 FPGAs in a node. Swift introduces a  decoupled, asynchronous model based on the Gather-Apply-Scatter (GAS) scheme. It subgraphs across FPGAs, and each subgraph into intervals based on source vertex IDs. Processing on these intervals is decoupled and executed asynchronously, instead of bulk-synchonous operation, where throughput is limited by the slowest task. This enables simultaneous processing  within each multi-FPGA node and optimizes the utilization of communication (PCIe), off-chip (HBM), and on-chip BRAM/URAM resources. Swift demonstrates significant performance improvements compared to prior scalable FPGA-based frameworks, performing 12.8 times better than the ForeGraph. Performance against Gunrock on NVIDIA A40 GPUs is mixed, because NVlink gives the GPU system a nearly 5X bandwidth advantage, but the FPGA system nevertheless achieves 2.6x greater energy efficiency.
%Furthermore, it achieves approximately 1.8 times better performance and 2.6 times greater energy efficiency compared to 
\end{abstract}


\section{Introduction}
The growth of graph data in fields such as social network analysis, biomedical research, and graph neural networks (GNNs)~\cite{ang_introducing_2010,lee_extrav_2017,noauthor_graph_2020, low_distributed_2012,zhou_graph_2019} has created a greater need for efficient and scalable graph analysis~\cite{mccune_thinking_2015, malewicz_pregel_2010, towards-dataflow-based-graph-processing, low_graphlab_2010, distributedgraphlab}. However, existing hardware platforms face limitations in handling large-scale graphs.  CPUs\cite{lakhotia_accelerating_2018,malewicz_pregel_2010,malicevic_everything_2017,beamer_gap_2017,shun_ligra_2013,zhu_gridgraph_2015} and GPUs~\cite{wang_gunrock_2016,zhang_gpu-accelerated_2011,khorasani_cusha_2014,xu_graph_2014} 
%provide fast and robust hardware capabilities for graph algorithms, they
often result in fine-grained random memory accesses~\cite{beamer_locality_2015, faldu_domain-specialized_2020,mughrabi_ecg_2024,balaji_p-opt_2021, basak_analysis_2019,pref}, which typically do not use memory bandwidth efficiently, limiting processing throughput.
%leading to poor utilization of the processing resources.

The emergence of modern FPGAs, with High Bandwidth Memory (HBM) technology (up to~\SI{460}{\giga\byte\per\second}) and custom parallel pipelines for data processing, presents an alternative  to address the challenges associated with accelerating graph processing~\cite{shi_exploiting_2022,jain_modular_2023,siefert_observed_2023,chen_regraph_2022,li_scalabfs2_2024,mughrabi_qpr_2021,hu_graphlily_2021,hashmem,jaiyeoba_acts_2023}. These state-of-the-art FPGAs offer significant parallelism compared to CPUs and GPUs, while maintaining a more efficient power profile. The inclusion of HBM allows graph data to be distributed across HBM channels and processed by specialized pipelines that achieve better memory access patterns, leveraging HBM's high memory bandwidth and enhancing parallelism. This feature makes HBM-enabled FPGAs a promising solution for graph processing. 

% 12.25Gb/s
% \SI{12.25}{\giga\bit/\second}
\begin{scriptsize}
    \begin{table*}[t]
        \centering
        \caption{PageRank (PR) - Comparing Single-FPGA vs. Multi-FPGA "Scale Out" Graph Accelerators from Prior Art}
        \scalebox{0.65}{
            \begin{tabular}{@{}lccccccccc@{}}
            \toprule 
            \toprule
            \textbf{Work} & \textbf{Lang.} & \textbf{Impl.}  & \textbf{Eval. Public} & \textbf{Platform} & \textbf{Mem (BW)} & \textbf{Interconnect/host (BW)} & \textbf{Throughput(GTEPS$^{\mathrm{a}}$)} & \textbf{FPGA(nodes)}  & \textbf{Reference} \\
            \midrule
            ReGraph   & HLS & HW      & \cmark & Alveo U280    & \SI{460}{\giga\byte/\second}   & \SI{38}{\giga\byte/\second} & 8.037$^{\mathrm{b}}$ & 1 & \cite{chen_regraph_2022} \\
            ThunderGP & HLS & HW      & \cmark & Alveo U250    & \SI{77}{\giga\byte/\second}   & \SI{38}{\giga\byte/\second} & 3.355$^{\mathrm{b}}$ &  1 & \cite{chen_thundergp_2021}\\
            GraphLily & HLS  & HW      & \cmark & Alveo U280    & \SI{460}{\giga\byte/\second}   & \SI{38}{\giga\byte/\second} & 5.591$^{\mathrm{b}}$ &  1 & \cite{hu_graphlily_2021}\\
            ACTS & HLS  & HW          & \cmark & Alveo U280    & \SI{460}{\giga\byte/\second}   & \SI{38}{\giga\byte/\second} & 8.557$^{\mathrm{b}}$ &  1 & \cite{jaiyeoba_acts_2023}\\
            \midrule
            ForeGraph      & HDL      & Sim      & -       & Xilinx VC707  & \SI{19.2}{\giga\byte/\second}  & \SI{98}{\giga\byte/\second} & 1.861, 3.675, 7.350, 9.8$^{\mathrm{d}}$  & 4, 8, 16, 32$^{\mathrm{e}}$ & \cite{dai_foregraph_2017}\\
            GraVF-M        & Python   & HW      & \cmark  & Microsemi     & \SI{21.7}{\giga\byte/\second}  & \SI{5.85}{\giga\byte/\second} & 4.623 & 4$^{\mathrm{e}}$  & \cite{engelhardt_gravf-m_2019} \\ 
            GridGAS        & HDL      & HW      & -       & Xilinx KC705  & -                              & \SI{3}{\giga\byte/\second} & 0.170$^{\mathrm{c}}$  & - & \cite{zou_gridgas_2018} \\ 
            FDGLib         & HDL  & HW/Sim      & \cmark  & Alveo U250    & \SI{77}{\giga\byte/\second}    & \SI{12.25}{\giga\bit/\second} & 2.679, 5.916, 18.816, 33.026 & 4, 8, 16, 32$^{\mathrm{e}}$ & \cite{wu_fdglib_2021}\\ 
            Hadoop         & HLS  & HW      & -       & Alveo U250    & \SI{19.2}{\giga\byte/\second}  & \SI{32}{\giga\byte/\second}  & 0.046 & 16$^{\mathrm{e}}$ & \cite{sahebi_distributed_2023}\\ 
            \midrule
            \midrule
            Swift          & HLS  & HW      & \cmark & Alveo U280    & \SI{460}{\giga\byte/\second}   & \SI{38}{\giga\byte/\second}  & 13.168,  22.407  & 4, 8$^{\mathrm{f}}$  & [current]\\ 
            \bottomrule 
            \bottomrule
             \multicolumn{9}{@{}p{\linewidth}@{}}{$^{\mathrm{a}}$ (Giga Traversed Edges Per Second) measures graph processing performance.} \\
             \multicolumn{9}{@{}p{\linewidth}@{}}{$^{\mathrm{b}}$ Geometric mean GTEPS$^{\mathrm{a}}$ was calculated for all graphs with sizes below 300 million edges, as reported in their papers, using a single FPGA setup.} \\
             \multicolumn{9}{@{}p{\linewidth}@{}}{$^{\mathrm{c}}$ Paper evaluation only for SSSP algorithm.} \\
             \multicolumn{9}{@{}p{\linewidth}@{}}{$^{\mathrm{d}}$ Paper peak performance for PR on Twitter Graph.} \\
             \multicolumn{9}{@{}p{\linewidth}@{}}{$^{\mathrm{e}}$ Single-FPGA per node - "scale out".} \\
             \multicolumn{9}{@{}p{\linewidth}@{}}{$^{\mathrm{f}}$ Multi-FPGA per node - "scale up".} \\
            \end{tabular}
        }
        \captionsetup{justification=centering}
        \label{table:single-versus-multi-fpga-accelerators}
    \end{table*}
\end{scriptsize}

\textbf{From single to multi-FPGA graph processing:} Prior research on single FPGA graph accelerators is comprehensive but typically supports graphs with fewer than 500 million edges~\cite{chen_regraph_2022,jaiyeoba_acts_2023,hu_graphlily_2021}. To improve scalability, multi-FPGA methods often use a single FPGA solution replicated across multiple machines or rely on inter-FPGA interconnects~\cite{zhao_lightgraph_2014,sahebi_distributed_2023,dai_foregraph_2017,wu_fdglib_2021,engelhardt_gravf-m_2019}, which can perform poorly in sharing fine-grained graph data. Such approaches can lead to inefficient processing, higher power consumption, increased memory usage, and complex inter-machine communication. Furthermore, scaling up a single-FPGA design within a machine is limited by the PCIe communication bandwidth.
%causing idle FPGA time during message exchanges.

Table~\ref{table:single-versus-multi-fpga-accelerators} presents a performance comparison of the widely-benchmarked PageRank algorithm among various single- and multi-FPGA graph processing frameworks. To enable fair comparison, metrics such as interconnect bandwidth and memory bandwidth of the FPGAs are also provided. As shown, multi-FPGA frameworks like Foregraph~\cite{dai_foregraph_2017} and FDGLib~\cite{wu_fdglib_2021} often have lower throughput than single-FPGA frameworks such as ThunderGP~\cite{chen_thundergp_2021, chen_regraph_2022} across various graph algorithms and workloads. This is because PCIe-connected FPGAs (multi-FPGA frameworks) exhibit lower latency compared to network-connected FPGAs (single-FPGA frameworks) due to fast/high bandwidth on-chip HBM memory coupled with PCIe DMA, providing direct access to host memory without the need to navigate the network stack \cite{futureoffpgaaccelerationindatacenters} \cite{scaleuporscaleout} \cite{scalingupterascalegraphprocessing}. This challenges the presumed superiority of multi-FPGA setups. ``Scaling up" refers to adding more compute power (FPGAs) to a single machine via PCIe, while ``scaling out" involves adding more machines with the same compute power via a specialized network—in this context, adding more machines with a single FPGA. % Our findings suggest better results might be achieved by scaling up single machines with multi-FPGA designs rather than scaling out.

\textbf{Scaling up single machine multi-FPGA graph processing and addressing communication overhead:} A key factor in the performance gap of multi-FPGA frameworks is the communication overhead among FPGAs. Frameworks like Foregraph~\cite{dai_foregraph_2017} necessitate costly inter-FPGA communication for exchanging vertex property information at each graph iteration due to the memory-bound nature of graph processing. FPGA memory bandwidth, such as with High Bandwidth Memory (HBM) at up to~\SI{460}{\giga\byte\per\second}, far exceeds that of inter-FPGA channels like PCIe at around~\SI{17}{\giga\byte\per\second}
%, exacerbating communication bottlenecks. Optimizing inter-FPGA communication is crucial for leveraging the potential of multi-FPGA setups for very large graphs, enabling them to surpass single-FPGA solutions.

\textbf{Swift: a decoupled Gather-Apply-Scatter graph execution model:} In order to address the communication bottleneck problem, we  decouple the main stages of the Gather-Apply-Scatter (GAS) graph processing scheme. This separation allows pipelining and overlapping the GAS compute and memory operations on the multi-FPGA system, enabling higher throughput while processing large graphs. With Swift, our decoupled graph processing model, four operations can run simultaneously: 1) processing edges at a given region (\textit{vertex intervals}) for a given iteration;
2) applying vertex updates to generate active frontiers at a second region.
3) exporting active vertices (\textit{frontiers}) to remote FPGAs in a third, different region. 
4) importing active frontiers from remote FPGAs in a fourth, different region. 
Swift's overlapping of GAS operations allows for higher utilization of available channels such as inter-FPGA communication channel (PCIe), intra-FPGA memory bandwidth (HBM), and on-chip BRAMs/URAMs. Furthermore, it improves throughput and conceals latency overheads. Swift adapts the open-source ACTS~\cite{jaiyeoba_acts_2023} FPGA accelerator for single-FPGA graph processing, and introduces {\em decoupled, asynchronous GAS processing} to overcome inter-FPGA communication latency and bandwidth limits, outperforming previous FPGA graph accelerator solutions. 

\section{Background and Related Work}

\begin{algorithm}[h]
\small % Reduces the font size
\caption{Edge-centric Gather-Apply-Scatter Model}
\label{alg:edge-centric-algorithm}
\SetAlgoLined
\KwData{Edges, vertices, and vertex properties}
\KwResult{Updated vertex properties ($V_{prop}$)}
\SetKwFunction{ProcessEdge}{Process\_Edge}
\SetKwFunction{Apply}{Apply}
\hrulefill\\
% \tcc{\textcolor{red}{Initialize variables}}

$E(U,V)$: Edge $E$, where $U$=source vertex ID, $V$=destination vertex ID \\
$E_{weight}$: Edge weight of the edge $E$ \\
$U(E)_{prop}$: Source vertex property\\
$V(E)_{prop}$: Destination vertex property\\
$V(E)_{temp\_prop}$: Temporary destination vertex property\\
$res$: Partial result (also known as \textit{vertex update}) generated from processing edge $E$ \\
\hrulefill\\
% \tcc{\textcolor{red}{Process each active streaming partition}}
\ForEach{active Streaming Partition $SP$ in graph}{
    \ForEach{outgoing edge $E(U,V)$ in $SP$}{
        \If{vertex $U$ is active}{
            $res \leftarrow$ \ProcessEdge($E_{weight}, U_{prop}$)
            % \tcp*{\textcolor{teal}{Process the edge to get partial result}}
            $V_{prop} \leftarrow$ \Apply($V_{temp\_prop}, res$) 
            % \tcp*{\textcolor{teal}{Apply the partial result to the destination vertex}}
        }
    }
}
\end{algorithm}

\begin{figure*}[t]
  \centering
  \includesvg[inkscapelatex=false,width=0.84\linewidth]{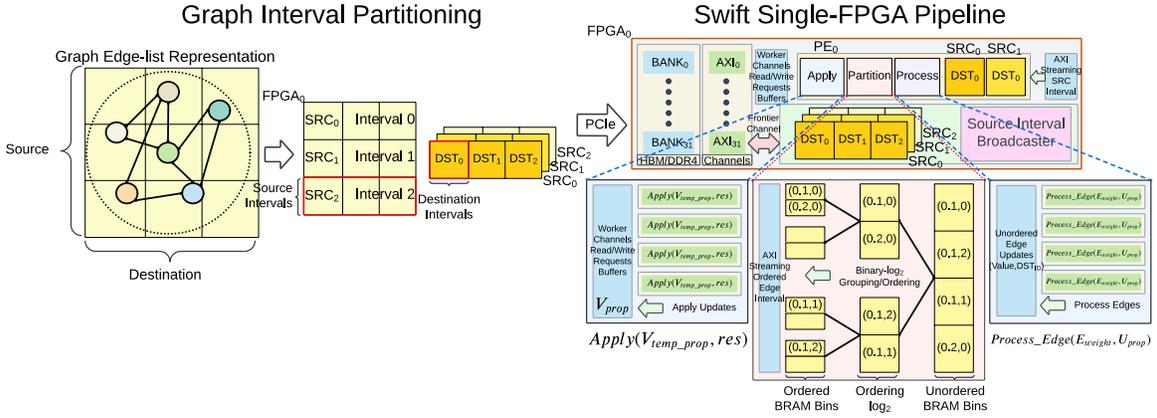}
  \caption{How Swift (ACTS~\cite{jaiyeoba_acts_2023} based pipeline) handles Process Edge, Partition-Updates, and Apply Update operations.}
  \label{fig:online-partitioning-in-swift} 
\end{figure*}

\subsection{Gather-Apply-Scatter (GAS)}
The Gather-Apply-Scatter (GAS)~\cite{gonzalez_powergraph_2012,gonzalez_graphx_2014,mccune_thinking_2015} model provides a high-level abstraction for various graph processing algorithms and is widely adopted by software-based~\cite{roy_x-stream_2013, giraph, low_graphlab_2010, malewicz_pregel_2010, sundaram_graphmat_2015} and accelerator-based frameworks~\cite{dai_foregraph_2017, chen_thundergp_2021, chen_regraph_2022, nurvitadhi_graphgen_2014, zhou_hitgraph_2019, ham_graphicionado_2016, zhou_high-throughput_2016}. The two main variants of the GAS model are the vertex-centric and edge-centric approaches. Swift adopts the edge-centric variant, which facilitates high throughput streaming memory accesses, leveraging HBM's high memory bandwidth. 

Algorithm \ref{alg:edge-centric-algorithm} shows the pseudo code describing the Edge-centric Gather-Apply-Scatter graph processing model~\cite{roy_x-stream_2013, zhu_gridgraph_2015}. As shown, this model employs streaming partitions by logically splitting the graph into intervals by source vertex IDs during pre-processing. Next, an input of an unordered set of directed edges is streamed and processed in the \textit{Process\_Edge} stage where edge data, source and destination vertex properties generate an update value (\textit{res}). Only intervals with active vertices are processed, avoiding redundant reads to all edges.  Furthermore, intervals are based on source IDs, and source vertex properties are read once from DRAM per iteration. In \textit{Apply}, these updates are applied to destination vertices to compute new vertex properties. These functions iterate until a convergence criterion is reached.

\subsection{ACTS: Near-Memory FPGA Graph Processing}
ACTS~\cite{jaiyeoba_acts_2023} is a graph processing accelerator that utilizes the edge-centric GAS model on FPGAs and employs HBM to address the memory bandwidth bottlenecks of prior single-FPGA-based graph processing designs. The key idea behind ACTS is an online recursive partitioning mechanism that converts (via partitioning) the low-locality vertex updates generated from processing the edges of an active sub-graph, into high-locality vertex-update partitions in efficient time. This partitioning is done across the {\em destination} vertex IDs. Through this, ACTS improves both read and write bandwidth performance, even as graph size increases. Consequently, ACTS achieved an average speedup of 1.5×, with a peak speedup of 4.6× compared to Gunrock~\cite{wang_gunrock_2016}, a state-of-the-art GPU-based graph processing accelerator, on the NVIDIA Titan X GPU. Furthermore, ACTS demonstrates an average speedup of 3.6×, with a peak speedup of 16.5× over GraphLily~\cite{hu_graphlily_2021}, a modern FPGA-based graph accelerator utilizing HBM. These speedups are found in their paper.  We therefore use this as the starting point for Swift's multi-FPGA solution.
\vspace{0px}

\subsection{ForeGraph: Scalable FPGA Graph Processing} ForeGraph~\cite{dai_foregraph_2017,wu_fdglib_2021} tackles scaling by using the Catapult torus interconnect in an FPGA simulated environment; however, it cannot scale beyond 48 nodes or maintain optimal performance as the number of nodes increases. As the number of FPGA nodes increases, the interconnet becomes a bottleneck, limiting scalability and degrading performance. %In many of these studies, the messages between the connections become a bottleneck between iterations, reducing performance as the number of FPGA nodes increases.
 %interconnects, such as the Catapult torus ForeGraph~\cite{dai_foregraph_2017, wu_fdglib_2021}, can become

\section{Swift}
\subsection{Graph Processing Decoupled Pipeline} \label{Graph Processing Operations}
The Swift graph processing accelerator builds upon ACTS by further decoupling its pipeline into five distinct stages: process-edge, partition-updates, apply-updates, import-frontier and export-frontier operations. This decoupling allows Swift to hide latency by exploiting overlap among operations, speeding up overall execution time. Figure~\ref{fig:online-partitioning-in-swift} illustrates the connection among three key stages—Process-edge, Partition-updates, and Apply-updates—involved in graph partitioning within an FPGA. Each stage interfaces with the HBM channels to receive specific data: edges for Process-edge, vertex updates for Partition-updates, and both vertex updates and properties for Apply-updates. The output data from each stage serves as the input for the subsequent stage, creating a continuous processing flow.

\begin{figure*}[t]
  \centering
  \includesvg[inkscapelatex=false,width=0.84\linewidth]{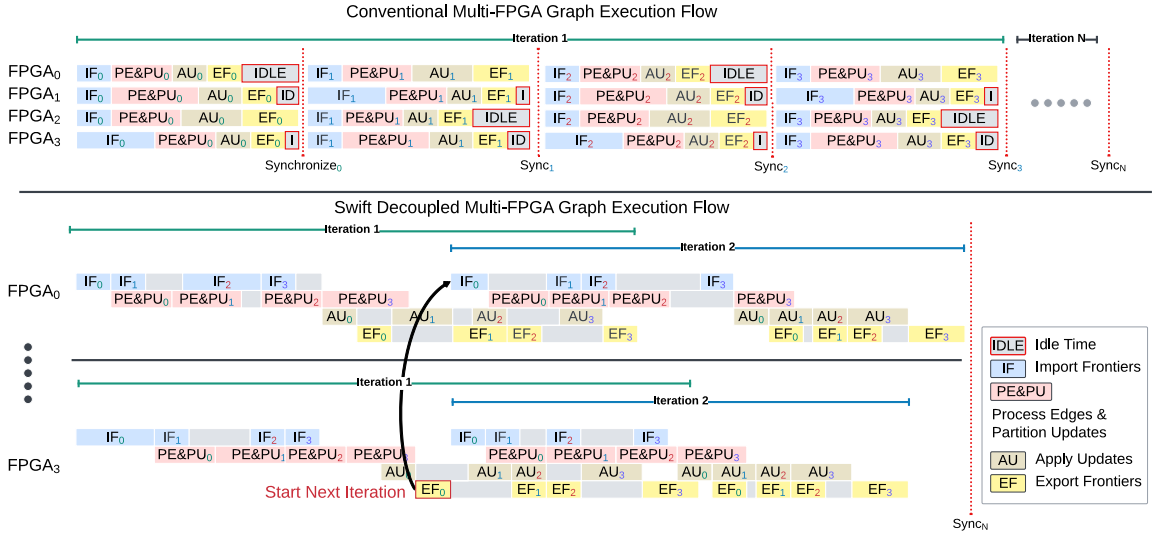}
  \caption{Swift decoupled multi-FPGA graph execution flow compared to prior art - decoupled operations execute asynchronously on the graph with no bulk synchronization.}
  \label{fig:decoupled-ops}
\end{figure*}

\begin{itemize}[topsep=1pt,itemsep=2pt,wide]
\item \textbf{Process-edge:} This operation generates vertex-update messages from the active sub-graph (source intervals). As shown in Figure~\ref{fig:online-partitioning-in-swift}, edges are read from HBM into \textit{EdgeProperty} Buffers (BRAM), and their source vertex properties are read into \textit{VertexProperty} Buffers (URAMs). Edges of active vertices are processed, and a user-defined edge function generates vertex-update messages, following the proposed scheme in Algorithm~\ref{alg:edge-centric-algorithm}. The vertex updates are buffered in DRAM for the partition-updates operation. The vertex-update tuple is formatted as (\textit{Value, Dst}), where \textit{Dst} is the destination vertex, and the value is the message. 

\item \textbf{Partition-updates:} 
The partition-updates stage, introduced in the ACTS paper~\cite{jaiyeoba_acts_2023}, addresses the challenge of random accesses and low spatial locality in vertex-updates generated from the process-edges stage. The operation is online and happens on the device side to further decompose the vertex updates generated from the process-edges. Partitioning enhances memory locality by converting low locality vertex updates (from edge processing) into high locality, enabling efficient use of fast URAMs for updating vertices. Due to the initial static partitioning of the graph, edge and vertex layouts within each HBM channel are optimized for online partitioning, which is confined to each HBM channel. The partition-updates operationconverts low-locality vertex updates into fine-grained, high-locality vertex-update partitions. The vertex updates generated from the process-edges operation are loaded into fast on-chip URAMs and BRAMs, and then partitioned using FPGA logic into high-locality partitions. This allows updates, represented by key-value pairs, to leverage the Ultra-RAM (URAM) multi-port parallelism and high capacity in Xilinx~\cite{noauthor_vitis_nodate} FPGAs when applied to destination vertices. However, with large graphs, URAM capacity is still limited, which can lead to partitioning overheads when swapping vertex updates. As shown in Figure~\ref{fig:online-partitioning-in-swift}, a recursive BRAM tree ($log_2(Dst)$) manages DRAM access latency with multilevel passes as updates move between BRAM and HBM, improving vertex-update locality with each level, thus reducing DRAM access latency. This makes it preferable to conventional bucket partitioning, especially for large graphs with low spatial locality. By breaking the task into recursive steps and buffering intermediate partial-partitioned results in HBM, the recursive BRAM tree strategy improves overall partitioning throughput and efficiency. The number of passes is the logarithm of the range of destination vertex IDs. This  ensures efficient data transfer from BRAM to HBM, surpassing  bucket partitioning. 

To illustrate the advantage of the recursive BRAM tree over conventional bucket-based partitioning, assume we have (N) vertex updates in HBM from processing active graph edges. Conventional bucket partitioning reads chunks of updates into the FPGA, splits them into (P) buckets based on destination vertex IDs, and writes them back to HBM. As the graph size increases, both (N) and (P) grow, requiring more partitions to maintain locality, which increases DRAM access latency and degrades performance. In contrast, the recursive BRAM tree strategy splits partitioning into successive steps, reducing latency and performance degradation. 
%While the benefits of the recursive approach may be less apparent for small graphs, they become significant as graph size increases. 
In each pass, after the buckets are filled, they are streamed into HBM, using its full bandwidth.  Then in the next pass, each bucket is read back from memory (also streaming), and partitioned again, until sufficient locality is achieved. Although it may require a logarithmic number of passes for large graphs, the recursive BRAM partitioning  allows for better performance than prior art, because it maintains high locality in the HBM accesses.

\item \textbf{Apply-updates:} 
When receiving a vertex update, the apply updates stage resolves this update to its destination vertex using a user-defined Apply function. This apply operation generates an active frontier property. Because the earlier (i.e., Partition-updates) operation outputs vertex-update chunks with high BRAM locality, the Apply operation can benefit from fast URAM memory. This is because several high-locality vertex-update partitions (generated from the partition-updates stage) and their corresponding destination vertex properties are streamed into independent high-speed URAMs, each connected to a separate apply-update logic. Therefore, several updates can be applied concurrently, allowing parallelism. In this way, the Apply operation can benefit from fast URAM memory.

\item \textbf{Import-frontier and export-frontier operations:}
A multi-FPGA graph processing context requires periodically exchanging graph data between FPGAs. Export-frontier operations send active frontiers from a given FPGA to its remote neighbors via PCIe through the host. These active frontiers are gathered and merged at the remote FPGA end using the import-frontier operation. Host-FPGA communication uses a host-managed shared buffer for DMA transfers over PCIe. This buffer moves active vertex properties (frontiers) between host memory and FPGA's HBM during export/import operations (Section III). The DMA engine handles memory transfers, reading from host memory (H2C) and writing to FPGA, and vice-versa.
\end{itemize}

\subsection{Understanding the Swift Pipeline} \label{Graph Execution Model}  
The Swift pipeline leverages the time window between edge processing within an FPGA and its next iteration to overlap with other intra-FPGA computation and inter-FPGA communication, using separate FPGA resources concurrently. Key to our flow model is that regions within the active sub-graph (vertex intervals) can start the next operation in the pipeline once dependencies are met. This contrasts to the bulk-synchronous model adopted by various prior art that require each operation to finish on the entire sub-graph before proceeding to the next. This allows for overlapping operations on two levels:
\begin{itemize}[topsep=1pt,itemsep=2pt,wide]
\item \textbf{Inter-FPGA:} Overlapping computation (within FPGAs) with communication operations (between FPGAs).%in the multi-FPGA machine.
\item \textbf{Intra-FPGA:} Operations within the same FPGA, hiding expensive, throughput-limiting operations within each other in the processing pipeline and improving throughput.
\end{itemize}

Figure~\ref{fig:decoupled-ops} shows Swift's decoupled execution flow versus the conventional bulk-synchronous model. In the conventional model, stages happen sequentially, starting only after the previous one is completed. For example, exporting active frontiers to remote FPGAs (export-frontiers stage) occurs only after applying updates to the active sub-graph (apply-updates stage). Similarly, processing edges (process-edges stage) occur only after receiving all import frontiers.
In Swift's model (Figure~\ref{fig:decoupled-ops}), a decoupled flow exploits potential overlaps within and between FPGAs. The graph is divided into partitions, each assigned to an FPGA. Within each FPGA, partitions are divided by source vertex IDs into vertex intervals, illustrated in Figure~\ref{fig:decoupled-ops} during pre-processing. Graph layout details are in Section~\ref{Workload Balancing}. For simplicity, four vertex intervals are shown. Each interval goes through five stages as in Section~\ref{Graph Execution Model}. Unlike the conventional model, there's no bulk-synchronous constraint. An interval can start its next operation as soon as its dependencies are satisfied. This is explained in Section~\ref{Example Execution Flow}.

To better understand the Swift decoupled pipeline, let us look at each overlapping feature when processing graphs:
\begin{itemize}[topsep=1pt,itemsep=2pt,wide]
\item \textbf{Overlap between computation within an FPGA and communication between FPGAs:} In Figure~\ref{fig:decoupled-ops}, $FPGA_{0}$ starts processing edges using the process\_edge operation (denoted by $PE_{0}$) in src interval 0 as soon as its active frontiers are imported from remote FPGAs. This happens concurrently with src interval 1 importing its active frontiers ($IF_{1}$). Similarly, Dst interval 0 in $FPGA_{0}$ exports active frontiers to remote FPGAs ($EF_{0}$), concurrently with interval 1 generating vertex updates ($AU_{1}$), allowing computation overlap within an FPGA and communication between FPGAs.
\item \textbf{Overlap between multiple operations within the same FPGA:} For example, import-frontier operation for interval 1 in iteration 2 runs concurrently with process\_edge and partition-updates for interval 0, and apply-updates for interval 3, overlapping operations within each FPGA and keeping HBM, URAM, and compute resources simultaneously busy.
\end{itemize}
Due to strict dependencies, some FPGA operations cannot overlap. Partition-updates and apply-updates are such operations. Apply-updates can begin only after vertex updates from process-edges are partitioned online. Additionally, process-edges and partition-updates can be merged into a single step.

\begin{figure*}[t]
  \centering
  \includesvg[inkscapelatex=false,width=0.84\linewidth]
  %\includesvg[inkscapelatex=false,width=0.70\linewidth]
  {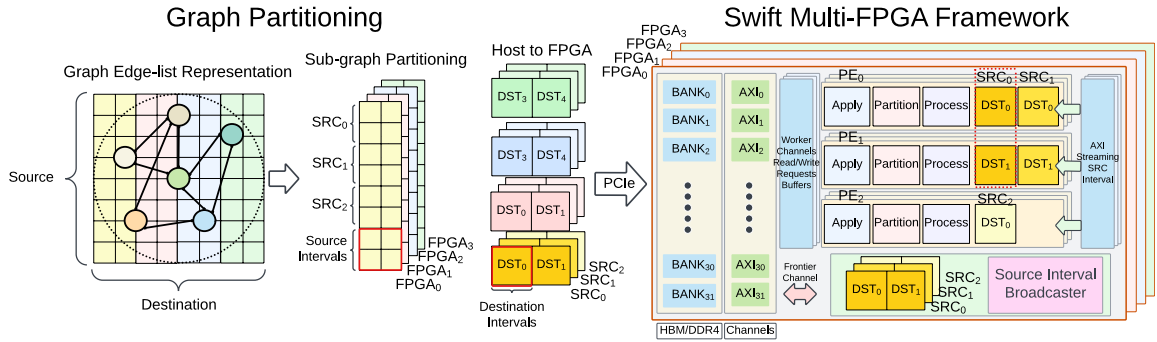}
  \caption{Swift framework ensures the graph is partitioned into load-balanced intervals distributed across FPGAs and Processing Elements (PEs), and then processed asynchronously.}
  \label{fig:graph-layout}
\end{figure*}

\subsection{Swift Flow Example} \label{Example Execution Flow}

This section will demonstrate how Swift operates within FPGA hardware, using a cluster of four FPGAs as an example. For simplicity, we will focus on the operations within a single FPGA ($FPGA_{0}$.) The workload graph assigned to each FPGA is first divided into vertex intervals based on vertex IDs, as shown in Figure~\ref{fig:decoupled-ops} and~\ref{fig:graph-layout}. A vertex interval consists of vertices along with their incoming edges. Inside each FPGA, five execution modules carry out the following operations: \textbf{Process-edges ($PE_M$)}, \textbf{Partition-updates ($PU_M$)}, \textbf{Apply-updates ($AU_M$)}, \textbf{Export-frontier ($EF_M$)}, and \textbf{Import-frontier ($IF_M$)}.

At any given time, each vertex interval can be in one of three states: "ready-for-process," "ready-for-export," and "ready-for-import." An interval in the "ready-for-process" state indicates that all dependencies needed for the process-edges operation on that interval have been met, allowing the process-edges module to execute that interval directly. The same principle applies to the "ready-for-export" and "ready-for-import" states. The modules continuously check the state of vertex intervals to carry out computation, export, and import operations. 
%The transitions between these states ensure consistency in the execution flow and the correctness of results.

The steps below demonstrate the Swift execution flow in FPGA hardware, focusing on one FPGA ($FPGA_{0}$). %within a four-FPGA cluster. 

\begin{enumerate}[topsep=1pt,itemsep=2pt,wide]
\item \textbf{Initialization:} During processing initiation, all vertex intervals containing active vertices in $FPGA_{0}$ to $FPGA_{3}$ are set to the ready-to-process state.

\item \textbf{Process-edges and Partition-updates:} The process-edge module ($PE_M$) in $FPGA_{0}$ activates in the ready-to-process state, executing process-edge and partition-update operations on all vertex intervals. Concurrently, the partition-updates module ($PU_M$) partitions low-locality vertex updates into high-locality vertex-update partitions.

\item \textbf{Apply-updates} After generating and partitioning vertex updates, the apply-updates module ($AU_M$) processes each partition to create active frontiers and flags the intervals as ready-for-export.

\item \textbf{Export-frontiers:} The export-frontier module ($EF_M$) starts exporting active frontiers to the host CPU when triggered by the ready-for-export flag. This enables overlap between apply-updates and export-frontier operations until all vertex interval frontiers are processed. Figure~\ref{fig:decoupled-ops} illustrates this with $AU_{0}$, $AU_{1}$, $AU_{2}$,  $AU_{3}$ overlapping $EF_{0}$, $EF_{1}$, $EF_{2}$,  $EF_{3}$.

\item \textbf{Import-frontiers:} Active frontiers associated with vertex intervals are marked ready-for-import by the export-frontier module. Thus, the import-frontier module ($IF_M$) in remote FPGAs can overlap their operations. This is shown by the overlap of $EF_{0}$, $EF_{1}$, $EF_{2}$, $EF_{3}$ from iteration 1 with $IF_{0}$, $IF_{1}$, $IF_{2}$, $IF_{3}$ from iteration 2 in Figure~\ref{fig:decoupled-ops}.
\item \textbf{Cycle Continuation:} This cycle continues until the algorithm converges (i.e., no more active frontiers) or until each vertex interval has completed a given number of  iterations.
\end{enumerate}

\section{Graph Partitioning and Workload Balancing} 
\subsection{Graph Partitioning} \label{Workload Placement}
\label{Workload Placement}
Figure \ref{fig:graph-layout} illustrates the layout of a graph within Swift FPGA cluster. The graph is initially partitioned by its destination vertex IDs across different FPGAs. Graph partitioning is performed on the host side as a pre-processing step, as represented by different colors. Since Swift is designed for static graphs (i.e., graphs with a fixed topology), this is treated as a one-time cost that can be amortized over multiple iterations. Each data type is partitioned differently. As in some prior work, vertex properties are represented using two dimensions: source/destination.  Each FPGA holds a full copy of the source vertex properties, while destination vertex properties are distributed across all HBM channels and all FPGAs, as are edges, which are partitioned by destination vertex IDs. Each processing element is connected to one HBM channel, processing edges and destination vertex properties in that channel. Each destination range and its incoming edges are assigned to a unique FPGA. Within each FPGA, the graph partition is further divided by source IDs. Each FPGA in the cluster has a dedicated HBM channel, the ``frontier HBM," which accommodates active frontiers imported from the communication channel. The remaining HBM channels in the FPGA, the ``worker HBMs," each store a segment of the graph's destination vertices and their incoming edges. Each processing element (PE) is linked to a worker HBM and handles the edges within that specific channel. To prevent graph data duplication and maintain storage efficiency, unique edges and vertices are distributed across  HBM channels.

The vertices within each HBM are categorized into vertex intervals, with the range being \( \frac{V}{NUM\_PEs} \), where $V$ represents the vertex properties that can fit in URAM. $NUM\_PEs$ denotes the number of processing elements in the cluster. Consequently, the combined range of vertex intervals across all PEs in the cluster is $V$.

\subsection{Workload Balancing} \label{Workload Balancing}
Optimizing performance in a cluster-scale environment with HBM-enabled FPGAs requires an efficient workload placement strategy that leverages parallelism at multiple levels. The first level of parallelism comes from the independent FPGAs in the cluster. Each FPGA has 32 independent HBM channels, adding a second level of parallelism. Swift configuration allows up to 128 Processing Elements (PEs) to operate independently in a 4-FPGA cluster. The challenge is to prevent any straggler PE from becoming a bottleneck, which requires a graph placement strategy that ensures uniform workload balance. Prior schemes distributes the graph across machines using a pre-processing step, maintaining balance but sacrificing throughput due to the graphs' unstructured nature \cite{foregraph, fdglib}. This resulted in the creation of cutting edges across various machines, disrupting the sequential ordering of vertex IDs. As a result, vertex translation was necessary for storage efficiency, and it also introduced communication bottlenecks between FPGAs. Swift avoids translations at receiver FPGAs by consistently referencing vertices using global IDs across all FPGAs. Additionally, it enforces a vertex-interval-based strategy for workload placement. This means that all vertices and edges within a vertex interval are placed across the entire cluster before moving to the next interval. This approach allows imported active frontiers to fit into low-latency URAM, enhancing throughput. In summary, our proposed strategy for placing graph workloads balances workload distribution and optimizes throughput in a cluster-scale, HBM-enabled FPGA environment. We achieve efficient graph processing without sacrificing overall performance by addressing translation bottlenecks and leveraging parallelism.

\section{Performance Evaluation}

\begin{scriptsize}
    \begin{table}[t]
        \centering
        \caption{Graph datasets under evaluation}
        \scalebox{0.75}{
            \begin{tabular}{@{}lrrrrr@{}}
                \toprule \toprule
                \textbf{Dataset} & \textbf{Symbol} & \textbf{\#Vertices} & \textbf{\#Edges} & \textbf{Type} \\
                \midrule
                Indochina & IND & $7.4$M & $194$M & Real \\
                Twitter & TW & $41.6$M & $1.4$B & Real \\
                Sk-2005 & SK & $50.6$M & $1.9$B & Real \\
                Uk-2005 & UK & $39.5$M & $936$M & Real \\
                Soc-sinaweibo & SN & $58.7$M & $523$M & Real \\
                Webbase-2001 & WB & $118$M & $1.0$B & Real \\
                RMAT\_8 & R8 & $8.39$M & $1.07$B & Syn \\
                RMAT\_16 & R16 & $16.8$M & $1.07$B & Syn \\
                RMAT\_32 & R32 & $33.6$M & $1.07$B & Syn \\
                \bottomrule \bottomrule
                % \multicolumn{6}{@{}p{\linewidth}@{}}{$^{\mathrm{a}}$  M: millions.} \\
                % \multicolumn{6}{@{}p{\linewidth}@{}}{$^{\mathrm{b}}$ B: billions.} \\
            \end{tabular}
        }
        \captionsetup{justification=centering}
        \label{datasets_for_evaluation}
    \end{table}
\end{scriptsize}

\begin{table*}[t]
    \centering
    \caption{Benchmark Tools and Hardware Specifications}
    \resizebox{0.95\linewidth}{!}{%
        \begin{tabular}{@{}lcccccccccccc@{}}
          
            %\toprule \toprule
            %\textbf{Benchmark} & \textbf{Devices per Node} & \textbf{Architecture} & \textbf{$\approx$BW$^{\mathrm{a}}$ (GB/s)} & \textbf{BW$^{\mathrm{b}}$ (GB/s)} & \textbf{BW$^{\mathrm{c}}$ (GB/s)} & \textbf{Runtime Power (W)} & \textbf{Freq$^{\mathrm{d}}$ (MHz)} & LUT & FF & BRAM & URAM \\
            
            %\midrule
            %Swift (FPGA)     & $\uparrow$ 4, 8 & Alveo U280           & 460 & 345 & 17 (PCIe)         & \SIrange{25}{50}{\watt}  & 150 MHz  & 870K (65.4\%) & 720K (25.4\%) & 2001 (49.3\%) & 768 (80.0\%)\\

            %Gunrock (GPU)~\cite{wang_gunrock_2016,noauthor_nvidia_nodate}    & $\uparrow$ 4  & Tesla A40            & 768 & 768 & 112 (NVLink)      & \SIrange{120}{180}{\watt} & 1305 MHz\\

            \toprule \toprule
            \textbf{Benchmark} & \textbf{Devices per Node} & \textbf{Architecture} & \textbf{$\approx$BW$^{\mathrm{a}}$ (GB/s)} & \textbf{BW$^{\mathrm{b}}$ (GB/s)} & \textbf{BW$^{\mathrm{c}}$ (GB/s)} & \textbf{Runtime Power (W)} & \textbf{Freq$^{\mathrm{d}}$ (MHz)} & LUT$^{\mathrm{e}}$ & FF$^{\mathrm{e}}$ & BRAM$^{\mathrm{e}}$ & URAM$^{\mathrm{e}}$ \\

            \midrule
            Swift (FPGA)     & $\uparrow$ 4, 8 & Alveo U280           & 1840 & 1380 & 68 (PCIe)         & \SIrange{100}{200}{\watt}  & 150 MHz  & 3480K (65.4\%) & 2880K (25.4\%) & 8004 (49.3\%) & 3072 (80.0\%)\\
            
            Gunrock (GPU)~\cite{wang_gunrock_2016,noauthor_nvidia_nodate}    & $\uparrow$ 4  & Tesla A40            & 3072 & 3072 & 448 (NVLink)      & \SIrange{480}{720}{\watt} & 1305 MHz\\

            \bottomrule \bottomrule
            \multicolumn{6}{@{}p{\linewidth}@{}}{$^{\mathrm{a}}$  Memory BW: Off-chip DDR4/HBM memory bandwidth for four (4) cards.} \\
            \multicolumn{6}{@{}p{\linewidth}@{}}{$^{\mathrm{b}}$  Total Effective BW: PCIe/NVLink bandwidth between the FPGA/GPU respectively.} \\
  
            \multicolumn{6}{@{}p{\linewidth}@{}}{$^{\mathrm{c}}$  Total Communication BW: Maximum bandwidth the algorithm can use upon deployment.} \\

            \multicolumn{6}{@{}p{\linewidth}@{}}{$^{\mathrm{d}}$  Max clock freq: Maximum on-chip clock frequency per card.} \\

            \multicolumn{6}{@{}p{\linewidth}@{}}{$^{\mathrm{e}}$ Total LUT, FF, BRAM and URAM utilization across 4 FPGAs} \\
            
            %\multicolumn{6}{@{}p{\linewidth}@{}}{$^{\mathrm{e}}$  Effective clock freq: Effective on-chip clock frequency per card.} \\
        \end{tabular}
    }
    \label{platform_specs}
\end{table*}

\begin{figure*}[h]
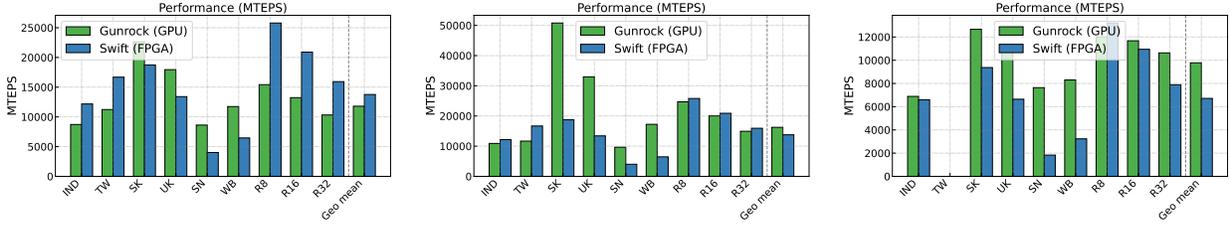

  \centering
  \begin{subfigure}[b]{.30\linewidth}
    \includesvg[inkscapelatex=false,width=\linewidth]{data/fig_pr_performance.svg}
    \label{fig:fig_performance_pr}
  \end{subfigure}
  \begin{subfigure}[b]{.30\linewidth}
    \includesvg[inkscapelatex=false,width=\linewidth]{data/fig_spmv_performance.svg}
    \label{fig:fig_performance_spmv}
  \end{subfigure}
   \begin{subfigure}[b]{.30\linewidth}
    \includesvg[inkscapelatex=false,width=\linewidth]{data/fig_hits_performance.svg}
    \label{fig:fig_performance_hits}
  \end{subfigure}
  \vspace*{-10pt}
  \caption{Performance Comparison of Gunrock (GPU) and Swift (FPGA) for PageRank (left), SpMV (middle), and HITS (right) using for 16 iterations (trials) 4 FPGAs/GPUs.}
  \label{fig:comparison_perf}
\end{figure*}

\begin{figure*}[h]
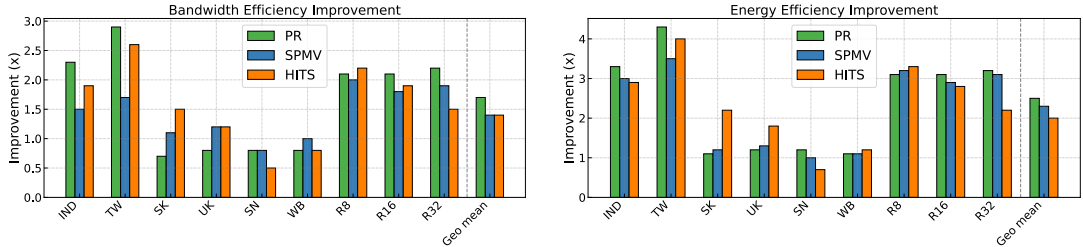

  \centering
  \begin{subfigure}[b]{.40\linewidth}
    \includesvg[inkscapelatex=false,width=\linewidth]{data/fig_bandwidth_improvement.svg}
    \label{fig:fig_bandwidth_improvement}
  \end{subfigure}
  \begin{subfigure}[b]{.40\linewidth}
    \includesvg[inkscapelatex=false,width=\linewidth]{data/fig_energy_improvement.svg}
    \label{fig:fig_energy_improvement}
  \end{subfigure}
  \vspace*{-10pt}
  \caption{Efficiency improvement for Swift over Gunrock --- PR, SPMV, and HITS.}
  \label{fig:performance_improvement}
\end{figure*}

\begin{figure*}[h]
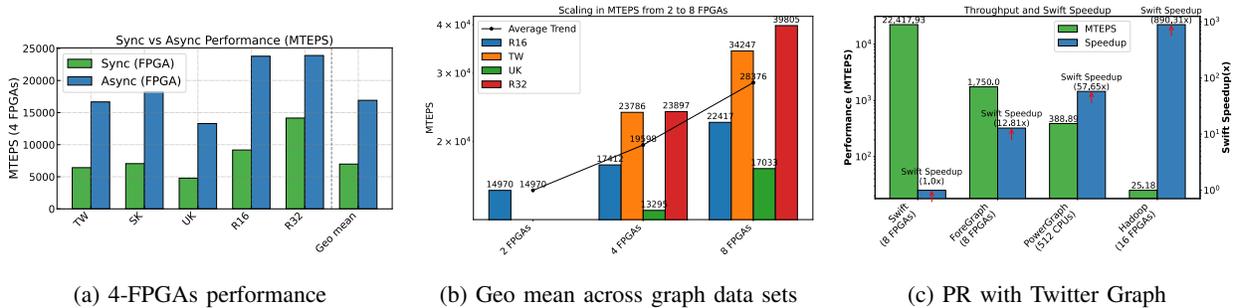

  \centering
  \begin{subfigure}[b]{.30\linewidth}
    \includesvg[inkscapelatex=false,width=\linewidth]{data/fig_sync_async_performance.svg}
    \caption{4-FPGAs performance}
    \label{fig:fig_async_improvement}
  \end{subfigure}
  \begin{subfigure}[b]{.30\linewidth}
    \includesvg[inkscapelatex=false,width=\linewidth]{data/fig_linear_scaling_3.svg}
    \caption{Geo mean across graph data sets}
    \label{fig:fig_scale_improvement}
  \end{subfigure}
  \begin{subfigure}[b]{.30\linewidth}
    \includesvg[inkscapelatex=false,width=\linewidth]{data/fig_performance_speedup_comparison_ex.svg}
    \caption{PR with Twitter Graph}
    \label{fig:comparison_fpgas}
  \end{subfigure}
  \caption{(a) Swift PageRank (PR) performance improvement (synchronous vs asynchronous) and (b) multi-FPGA scalability. (c)  shows Swift MTEPS and speedup vs.\ ForeGraph~\cite{dai_foregraph_2017}, PowerGraph~\cite{gonzalez_powergraph_2012}, and Hadoop~\cite{sahebi_distributed_2023} on Twitter Graph.}
  \label{fig:performance_async_scaling}
\end{figure*}

\subsection{Experimental Methodology}
\subsubsection{Graph Algorithms and Datasets}
We studied three commonly used graph algorithms: Pagerank (PR), Sparse Matrix-Vector Multiplication (SpMV), and Hyperlink-Induced Topic Search (HITS) to explore their distinct contributions within Swift. These algorithms capture memory access patterns that are common to various other graph algorithms. Our experiments involved using both synthetic and real-world datasets as shown in Table~\ref{datasets_for_evaluation}. We choose these datasets because they express diverse cache behaviors. The synthetic datasets were generated from the RMAT graph generator \cite{rmat-generator-library}, while the real-world datasets were obtained from the University of Florida’s Sparse Matrix Collection \cite{sparsesuite}. Because of the limited HBM memory capacity (8GB per FPGA), we have postponed the exploration of very large graphs for future research. Future HBMs are expected to deliver up to 32GB, allowing for much larger graphs to be run.

% \textcolor{red}{The system environment for Swift and Gunrock for each FPGA/GPU is shown in} Table~\ref{table:system2})
\subsubsection{Acceleration Environment, Design, Baselines, and Performance Metrics}
In our study, we conducted a comprehensive comparison of Swift with several state-of-the-art clusterscale systems, including ForeGraph (FPGA-based), PowerGraph (CPU-based), TurboGraph (FPGA-based), FPGP (FPGA-based), FDGLib (FPGA-based), and Gunrock (GPU-based). We compared Swift with Gunrock, as it was open-sourced. The complete implementation of Swift, including I/O and FPGA kernel invocation costs, was carried out using four Xilinx Alveo Ultrascale+ FPGA Accelerator Cards. These cards are equipped with HBM (High Bandwidth Memory) capable of delivering up to 460GB/s per FPGA. Gunrock, on the other hand, was tested on four NVIDIA A40 GPUs with HBM2 memory supporting 696GB/s per GPU (as shown in Table~\ref{platform_specs}). Communication among FPGAs in Swift occurs via PCIe, with data routed through the host. Our model uses PCIe's duplex feature for simultaneous read/write to optimize transfer, using Gen3 x16 PCIe, which delivers up to 17 GB/s. The RTL code was generated from the C++ HLS source using the Xilinx HLS tool, and the design was synthesized and run on the Xilinx Alveo FPGA board using the Xilinx Vitis tool. It is important to note that Vitis was only able to synthesize up to 24 Processing Elements (PEs), resulting in a clock frequency of 150 MHz. Further improvements to the synthesis, to enable more PEs and a higher clock frequency, are ongoing work, but this configuration already shows the potential of Swift. Timing measurements for both Swift and Gunrock begin once the graph is loaded onto the accelerator and end upon completion of kernel processing, capturing all exchanges during execution. The graph loading time is excluded, as it is a one-time cost. 
%Gunrock’s source code clearly separates data loading from kernel execution, and its authors focus on kernel execution times, making it directly comparable with our approach}.

%The host-side code includes modules for loading the kernel and graph data onto the FPGA (kernel loading code) and for synchronizing frontier property data among FPGAs during export/import operations (data synchronization). The kernel loading code is responsible for loading the kernel binary onto the FPGA device and signaling the host when the load is complete. The data synchronization code manages data exchanges between FPGAs, coordinating the movement of active vertex properties (i.e., active frontiers) between FPGA and host during export, and between host and FPGA during import. This 
Data movement is facilitated by the host using the DMA engine, which reads graph data from the host’s allocated memory (for H2C) and writes it directly into the FPGA’s HBM memory during import, and vice versa for export. The synchronization process is overlapped with the FPGA kernels performing graph processing across multiple FPGAs. This overlap is achieved through double buffering, out-of-order command queuing, asynchronous event handling, and non-blocking calls, all implemented on the host side.

\subsection{Results} 
\subsubsection{Throughput}
Figure~\ref{fig:comparison_perf}
% ,~\ref{fig:comparison_spmv},~\ref{fig:comparison_hits} 
and~\ref{fig:performance_improvement} present a comparative analysis of Swift with prior accelerators, leading to several noteworthy observations.

\begin{itemize}
\item Swift exhibits mixed performance compared to Gunrock in Figure \ref{fig:comparison_perf}.  Some datasets such as SK-2005 and UK-2005 are characterized by high regularity and cache hit rates, and have significant benefits from the advanced caching mechanism of the GPU. In contrast, Swift demonstrates superior throughput with relatively unstructured datasets over Gunrock. We could not collect results for HITS on Swift due to out-of-memory error.

\item It's important to note that the evaluation of Gunrock is on a GPU cluster using A40 GPUs with NVlink; the A40 offers  higher off-chip memory bandwidth (768 GB/s vs. 345 GB/s) and NVlink offers higher inter-device bandwidth (112 GB/s vs. 17 GB/s) compared to the Alveo U280 FPGAs. To evaluate the benefit the GPU system derives from NVLink vs. the slower PCIe interconenct, we evaluated Gunrock on PageRank on R8 with NVlink disabled. Without NVlink, Gunrock is 4.8X slower than with NVlink, similar to the bandwidth difference. {\em This suggests that with a similar high-speed interconnect, the multi-FPGA system would consistently outperform Gunrock for all our algorithms and datasets.}
% (Unfortunately, we ran out of time to gather complete non-NVlink results; we would include complete results in the final paper.) 

\item Swift exhibits superior performance over prior multi-FPGA-based (Foregraph, Hadoop with FPGAs) and CPU-based (Powergraph) clusterscale graph accelerators. As shown in Figure \ref{fig:comparison_fpgas}, Swift outperforms Foregraph by up to 12x, Hadoop by 890x, and Powergraph by 57x. This superiority can be attributed to two main factors:
    \begin{itemize}
    \item Swift effectively manages random accesses related to vertex-to-vertex communication within each FPGA by restructuring vertex updates during processing and leveraging fast URAMs to perform apply-update operations (refer to section \ref{Graph Processing Operations})
    
    \item Swift's decoupling strategy enables tight interleaving between computation (within FPGAs) and communication (between FPGAs), as well as between computation operations within the same FPGA. This reduces idle times.
    \end{itemize}

\item We compared Swift against the bulk-synchronous GAS approach (where no overlapping exists) to gain insights into the impact of our decoupling approach and better quantify the performance impact of overlapping communication with computation during graph processing. The result are plotted in Figure \ref{fig:fig_async_improvement}. To achieve this we turned off the asynchronous behavior in Swift to enforce that each iteration completes a bulk-synchronous step before the next commences. The results prove that Swift's decoupling mechanism provides about 2-3X improvement to throughput.
\end{itemize}

\subsubsection{Energy \& Bandwidth Efficiency}
Swift (FPGA-based) and Gunrock (GPU-based) were run on different platforms with different characteristics. The Alveo U280 FPGA has an off-chip memory bandwidth of 460GB/s, while the Tesla A40 GPU supports up to 768GB/s. To compare, we use bandwidth efficiency (MTEPS/bandwidth), and energy efficiency (MTEPS/Watt)
%We conducted experiments to measure the bandwidth and energy efficiency of both accelerators for a fair comparison. The Bandwidth efficiency (MTEPS/Bandwidth) is calculated by dividing the throughput (MTEPS) of each dataset by the effective memory bandwidth of the platform (as shown in Table \ref{platform_specs}, \textit{BW$^{\mathrm{b}}$ (GB/s)}) while the Energy efficiency (MTEPS/Watt) is calculated by dividing the throughput (MTEPS) of each dataset by its corresponding power consumption. \hl{We measured power consumption in Watts. 
We query GPU power using Nvidia-smi and FPGA using Xilinx’s xbutil. 
% Figure \ref{fig:fig_bandwidth_pr}, \ref{fig:fig_bandwidth_spmv} and \ref{fig:fig_bandwidth_hits} show the bandwidth efficiency results with improvement summary in Figure \ref{fig:fig_bandwidth_improvement} while Figure \ref{fig:fig_energy_pr}, \ref{fig:fig_energy_spmv} and \ref{fig:fig_energy_hits} show the energy efficiency results with improvement summary in Figure \ref{fig:fig_energy_improvement}. 
Based on our observations, Swift experiences about 1.5X better bandwidth efficiency than Gunrock and about 2X better energy efficiency. Further profiling of power consumption in Swift revealed that as much as 80\% of Swift's overall power  is used by the HBM  while only about 20\% is spent in on-chip FPGA activity. 
\vspace{0px}

\subsubsection{Scalability}
Figure \ref{fig:fig_scale_improvement} shows the throughput for a number of datasets plotted across an increasing number of FPGAs, to gain insights into how Swift scales. Some datasets (TW, UK \& R32) were too large to fit in a 2-FPGA setup and their 2-FPGA numbers were omitted. As shown, Swift's throughput increases relatively linearly as more FPGAs are added. This linear stability facilitated by the workload balancing mechanism  (Section \ref{Workload Balancing})  allows a graph workload to be uniformly distributed across the different FPGAs in the cluster.

\section{Conclusions} 
The paper introduces Swift, a clusterscale graph accelerator for FPGAs with HBM. Swift leverages the open-source ACTS~\cite{jaiyeoba_acts_2023} framework and addresses key challenges not present in single-FPGA accelerators, in particular the limited bandwidth of FPGA-to-FPGA communication and inefficiency in prior workload balancing strategies. To overcome these challenges, Swift %proposes a decoupled, asynchronous execution model based on the Gather-Apply-Scatter (GAS) paradigm. This model 
allows overlapping of crucial graph processing primitives, such as edge processing within a local FPGA, importing of active frontiers from remote FPGAs, and exporting of active frontiers to remote FPGAs. This approach maximizes communication bandwidth across PCIe, off-chip (HBM/DDR), and on-chip (SRAM), effectively concealing inter-FPGA communication with intra-FPGA computation. 
%Additionally, the paper imposes constraints on workload placement strategies to maintain a high throughput exchange datapath in the cluster. 
Swift outperforms prior FPGA-based frameworks. Results compared to Gunrock on a multi-GPU system are mixed, because the GPU system benefits from 5X higher inter-card bandwidth due to NVlink, but still achieves over 2X greater energy efficiency. If the FPGA system had a similar high-bandwidth interconnect, it should consistently outperform the GPU.
%, with much better energy efficiency.

\subsection*{Acknowledgements} %%% Added by KS after the paper was accepted
{\small This work was funded in part by PRISM, one of seven centers in JUMP 2.0, an SRC program sponsored by DARPA; the NSF I/UCRC MIST Center, 
%under grants IIP-1439644, IIP-1439680, IIP-1738752, IIP-1939009, IIP-1939050, and IIP-1939012; 
and Booz Allen Hamilton under contract FA-8075-18-D-0004. We also thank the anonymous reviewers for their helpful suggestions.}

%%%%%%%%% -- BIB STYLE AND FILE -- %%%%%%%%
\bibliographystyle{IEEEtran}
\bibliography{references,references_z}
%%%%%%%%%%%%%%%%%%%%%%%%%%%%%%%%%%%%

\end{document}